\documentclass[journal]{IEEEtran}
\usepackage{cite}

\ifCLASSINFOpdf
   \usepackage[pdftex]{graphicx}
\else
   \usepackage[dvips]{graphicx}
\fi
\usepackage{amsmath}
\usepackage{algorithmic}

\usepackage{array}

\ifCLASSOPTIONcompsoc
  \usepackage[caption=false,font=normalsize,labelfont=sf,textfont=sf]{subfig}
\else
  \usepackage[caption=false,font=footnotesize]{subfig}
\fi
\usepackage{dblfloatfix}
\usepackage{url}

\usepackage{bbm}
\usepackage{amssymb}
\usepackage{amscd}
\usepackage[amsmath,thref,thmmarks]{ntheorem}
\theoremstyle{plain}

\newtheorem{definition}{Definition}

\usepackage{hyperref}

\begin{document}
%
\title{CSI Learning Based Active Secure Coding Scheme For Detectable Wiretap Channel}
%
%
%

\author{Yizhi~Zhao

}

\maketitle

\begin{abstract}
In this paper, we consider the problem of secure and reliable communication with uncertain channel state information (CSI) and present a new solution named active secure coding which combines the machine learning methods with the traditional physical layer secure coding scheme. First, we build a detectable wiretap channel model by combining the hidden Markov model with the compound wiretap channel model, in which the varying of channel block CSI is a Markov process and the detected information is a stochastic emission from the current CSI. Next, we present a CSI learning scheme to learn the CSI from the detected information by the Baum-Welch and Viterbi algorithms. Then we construct explicit secure polar codes based on the learned CSI, and combine it with the CSI learning scheme to form the active secure polar coding scheme. Simulation results show that an acceptable level of reliability and security can be achieved by the proposed active secure polar coding scheme.
\end{abstract}

\begin{IEEEkeywords}
active wiretap channel, active secure coding, eavesdropper behavior learning, polar code, hidden Markov model.
\end{IEEEkeywords}

%
\IEEEpeerreviewmaketitle

\section{Introduction}
%
%
%
%
\IEEEPARstart{O}{ver} the last decade, polar code\cite{Arikan2009} based physical layer secure coding schemes\cite{Mahdavifar2011,Vard2013strong,Hassani2014,Gulcu2015,Wei2015,Zheng2017} have achieved secure and reliable communication over the wiretap channels (WTCs)\cite{Wyner1975} with a decisive assumption that legitimate parties perfectly know the precise channel information.
However in practical situation, uncertainties of the channel information always exists on legitimate side\cite{Ozarow1984,Wang2013,Schaefer2015,Goldfeld2016}. For instance, eavesdropper can initially choosing the wiretapping channels and keep the channel information unknown to legitimate parties. Such uncertainties of channel information have brought enormous limitations on the application of physical layer secure coding in realistic communication.

To solve this uncertain CSI problem, encryption methods are employed and combined with the physical layer secure coding\cite{Cheng2012,Nafea2016,Wang2013}, which however brings new limitations. Since the security of encryption is negatively related to the level of eavesdropper's computational power, the encryption-coding solution cannot be applied in some extremely high computational power cases, which is problematic due to the rapidly developing of quantum computing\cite{Shor1994,Monz2016} and the inevitable needs of anti-quantum computing communication. Therefore, we need to investigate new cooperative methods, other than the encryption, with the physical layer secure coding for the uncertain CSI problem.

\subsection{Our Work}

In this work we study a new solution for the uncertain CSI problem. In \cite{Tahmasbi2017} a \emph{detectable assumption} was proposed that \emph{modifications of eavesdropper's action may induce physical effects in the environment that can be detected by legitimate parties}. Based on this assumption, we have a general idea of active solution that if legitimate parties can learn the behavior of eavesdropper from the detected information and decode the current CSI, then the physical layer secure codes can be adjusted accordingly and actively.

To implement this idea, we carry out our work from three major aspects. First is the construction of a new WTC model that covers both uncertain CSI and detectable assumption. Second is the construction of a proper scheme to analyse the CSI from detected information. Third is the construction of active secure coding scheme that combines the CSI analysing scheme with the physical layer secure coding scheme.

Our contributions are summarized as follow:
\begin{itemize}
  \item[1)] We have built a new detectable assumption based WTC model by combining the hidden Markov model\cite{Welch2003} with the compound WTC model (a block varying WTC model\cite{Goldfeld2016}). In this new model, we use Markov process to express the varying of channel block CSI, and use a stochastic emission from the current CSI to generate the detected information.
  \item[2)] We have presented a CSI learning scheme to analyse the CSI from the detected information. Specifically, we setup a pre-collecting stage to collect the training data prior to the secure communication; we construct a CSI pattern learning scheme to learn the hidden Markov model from the detected information by the Baum-Welch algorithm \cite{Welch2003}; we also construct a CSI decoding scheme to decode the CSI from the detected information by the Viterbi algorithm\cite{Van2007} with the learned hidden Markov model.
  \item[3)] We construct an explicit secure polar codes based on the learned CSI, and combine it with the CSI learning scheme to form the active secure polar coding scheme. We also analyse the performance of the CSI learning scheme and the active secure polar coding scheme. For the analysis results of CSI learning scheme, the CSI pattern learning can achieve very good accuracy, but the CSI error rate of CSI decoding is not vanishing. Then for the analysis results of active secure coding scheme, both legitimate bit error rate and the information leakage rate stay at a low level, thus acceptable reliability and security can be achieved.
\end{itemize}

\subsection{Related Works}

The detectable assumption was proposed in \cite{Tahmasbi2017} and further studied in \cite{Tahmasbi2019}. In these study, the authors present a detectable WTC model and a corresponding secure coding scheme. But different from our model that legitimate can only observe CSI related information, \cite{Tahmasbi2017} assumes that legitimate parties directly obtain the CSI from the detected information with hindsight. Then with this delayed CSI, \cite{Tahmasbi2017} presents a encryption based secure codes.

Another WTC model that relates to our hidden Markov based detectable WTC model is the finite state Markov WTC with delayed feedback studied in \cite{Dai2017}. In this work, the authors have characterized the capacity-equivocation regions of their proposed model. Same with our model, the varying process of the CSI is a stochastic Markov process. What different is that in our model legitimate parties can only detect information relevant to the CSI but in \cite{Dai2017}'s model the legitimate receiver directly knows the CSI and can transmit it back to the legitimate sender as a delayed feedback.


\subsection{Paper Organization}

The outline of this paper is as follow. Section~\ref{sec_HMMwtc} presents the construction of the hidden Markov based detectable WTC model. Section~\ref{sec_active_coding} presents the construction of CSI learning scheme and the active secure polar coding scheme. Section~\ref{sec_simulation} presents the performance analysis of the active secure polar coding scheme. Finally, Section~\ref{sec_con} concludes the paper.

\subsection{Notations}

We define integer interval $[\![a,b]\!]$ as the integer set between $\lfloor a\rfloor$ and $\lceil b\rceil$. For $n\in \mathbb{N}$, define $N\triangleq 2^n$. Denote $X$, $Y$, $Z$,... random variables (RVs) taking values in alphabets $\mathcal{X}$, $\mathcal{Y}$, $\mathcal{Z}$,... and the realization of these RVs are denoted by $x$, $y$, $z$,... respectively. Then $p_{XY}$ denotes the joint probability of $X$ and $Y$, and $p_X$, $p_Y$ denotes the marginal probabilities. Also we denote a $N$ size vector $X^N\triangleq (X_1,X_2,...,X_N)$, denote $X_a^b\triangleq(X_a,X_{a+1},...,X_b)$. And for any index set $\mathcal{A}\subseteqq [\![1,N]\!]$, we define $X^\mathcal{A}\triangleq \{X_i\}_{i\in \mathcal{A}}$. For the polar codes, we denote $\mathbf{G}_N$ the generator matrix , $\mathbf{R}$ the bit reverse matrix, $\mathbf{F}=
    \begin{bmatrix}\begin{smallmatrix}
        1 & 0 \\
        1 & 1
    \end{smallmatrix}\end{bmatrix}$
, $\otimes$ the Kronecker product, and have $\mathbf{G}_N=\mathbf{RF}^{\otimes n}$. $H(\cdot)$ denotes the binary entropy and $I(\cdot)$ denotes the mutual information.

\section{Hidden Markov Based Detectable WTC Model}\label{sec_HMMwtc}

In this section, we build a new uncertain CSI WTC model with the detectable assumption.

\subsection{Hidden Markov Model}

In the detectable uncertain CSI case, CSI of the model is time varying and controlled by the eavesdropper. Legitimate parties only have the detected information relevant to the current CSI. To build such a model, there are two aspects for consideration.
\begin{itemize}
  \item \emph{CSI pattern:} a model that generates the time varying process of CSI, written as $\mathbb{P}$.
  \item \emph{Mapping of detected information:} a model that generates the detected information from the CSI, written as $\mathbb{O}$.
\end{itemize}

One good model that can match both CSI pattern and mapping of detected information is the \emph{hidden Markov model (HMM)}\cite{Welch2003}.

\begin{figure}[!h]
\centering
\includegraphics[width=8.5cm]{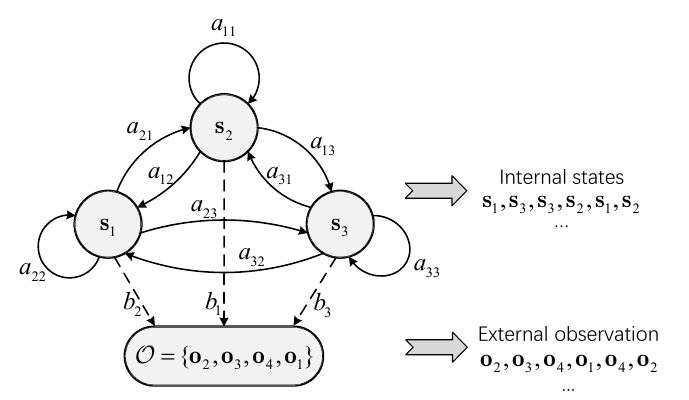}
\caption{The Hidden Markov Model.}
\label{fig_hmm}
\end{figure}

The structure of a basic HMM $(\mathcal{S},\mathcal{O},\mathbf{A},\mathbf{B},\pi)$ is illustrated in Fig.~\ref{fig_hmm} which contains following two parts:
\begin{itemize}
  \item \emph{A hidden internal part:} this part is a stochastic Markov process of hidden states which cannot be observed. As illustrated in Fig.~\ref{fig_hmm}, we have the Markov process as $(\mathcal{S},\mathbf{A},\pi)$. $\mathcal{S}=\{\mathbf{s}_1,\mathbf{s}_2,\mathbf{s}_3\}$ is the alphabet of possible state. $\mathbf{A}$ is a $3\times3$ state transparent probability matrix that each item $a_{ij}$ represents the probability of state changing from $\mathbf{s}_i$ to $\mathbf{s}_j$. $\pi$ is the probability matrix for the initial state $S_1$. For the beginning of this Markov process, the value of initial state $S_0$ is distributed according to $\pi$ over the alphabet $\mathcal{S}$. Then for any $t>1$, each state $S_t$ is stochastically determined by the last state $S_{t-1}$ and the state transparent probability matrix $\mathbf{A}$. For instance, if current state value is $\mathbf{s}_1$, then the next state value could be $\mathbf{s}_1$ with probability $a_{11}$ or $\mathbf{s}_2$ with probability $a_{12}$ or $\mathbf{s}_3$ with probability $a_{13}$.

  \item \emph{An external emission part:} this part is a stochastic mapping from the hidden state to the observable information. As illustrated in Fig.~\ref{fig_hmm}, we have this stochastic mapping as $(\mathcal{S},\mathcal{O},\mathbf{B})$. $\mathcal{O}=\{\mathbf{o}_1,\mathbf{o}_2,\mathbf{o}_3,\mathbf{o}_4\}$ is the observation alphabet. $\mathbf{B}$ is a $3\times4$ emission probability matrix that each item $b_{ij}$ represents the probability of mapping $\mathbf{s}_i$ to $\mathbf{o}_j$. For any time $t$, the value of observed information $O_t$ is stochastically determined by the value of current hidden state $S_t$ and the emission probability matrix $\mathbf{B}$. For example, if current state value is $\mathbf{s}_1$, then the value of observed information could be $\mathbf{o}_1$ with probability $b_{11}$ or $\mathbf{o}_2$ with probability $b_{12}$ or $\mathbf{o}_3$ with probability $b_{13}$ or $\mathbf{o}_4$ with probability $b_{14}$.

\end{itemize}

The two parts of HMM can match our modeling needs. The internal part can be used as the model for CSI pattern and the external part can be used as the model for detected information. Therefore, we build the detectable WTC model based on the HMM.

\subsection{HMM Based Detectable WTC Model}

First we make a few assumptions for the detectable WTC model.
\begin{itemize}
  \item Both main channel and wiretap channel are block varying that the CSI remains constant within each $N$ length block.
  \item Both main channel and wiretap channel are not necessarily symmetric or degraded.
  \item The varying and detecting operation of CSI is ahead of the encoding process, so legitimate parties can have the detected information of current CSI before the encoding.
  \item Legitimate parties know all the possible values of CSI and detected information.
\end{itemize}

Now we present the definition of our HMM based detectable WTC model.

\begin{definition}\label{def_hmm_wtc}
The HMM based detectable WTC model is defined as $(\mathcal{X},\mathcal{Y},\mathcal{Z},\mathcal{S},\mathcal{O},\mathbf{A},\mathbf{B},\pi)$ which contains an HMM $(\mathcal{S},\mathcal{O},\mathbf{A},\mathbf{B},\pi)$ with parameter set $\lambda_{\mathrm{H}}=(\mathbf{A},\mathbf{B},\pi)$.

$\mathcal{X}$ is the alphabet for main channel input, $\mathcal{Y}$ is the alphabet for main channel output, $\mathcal{Z}$ is the alphabet for wiretap channel output.

$\mathcal{S}=\{\mathbf{s}_1,\mathbf{s}_2,...,\mathbf{s}_\alpha\}$ is the finite alphabet of the CSI state also as the state set $\mathcal{S}$ of HMM, have $|\mathcal{S}|=\alpha$. For $\mathbf{s}_i\in \mathcal{S}$, $\mathbf{s}_i=p_{YZ|X}^{(i)}$ which have:
\begin{equation}
\begin{split}
&\forall\left(x^N,y^N,z^N\right)\in \mathcal{X}^N\times\mathcal{Y}^N \times \mathcal{Z}^N,\\
&p_{Y^NZ^N|X^N}\left(y^N,z^N|x^N\right)=\prod_{j=1}^{N}p_{YZ|X}^{(i)}\left(y_j,z_j|x_j\right),
\label{eq_channeldefine}
\end{split}
\end{equation}
and both main channel and wiretap channel are asymmetric with no degradation relationship.

$\mathcal{O}=\{\mathbf{o}_1,\mathbf{o}_2,...,\mathbf{o}_\gamma\}$ is the finite alphabet of detected information also as the observation set $\mathcal{O}$ of HMM, have $|\mathcal{O}|=\gamma$;
Matrix $\mathbf{A}$ is a $\alpha\times\alpha$ state transparent probability matrix
of the internal Markov process of HMM, for $i\in [\![1,\alpha]\!]$ have
\begin{equation}
\mathbf{A}=(a_{ij})_{\alpha\times\alpha},~a_{ij}=p(\mathbf{s}_j|\mathbf{s}_i).
\end{equation}

Matrix $\mathbf{B}$ is a $\alpha\times\gamma$ emission probability matrix of HMM from $\mathcal{S}$ to $\mathcal{O}$, for $i\in [\![1,\alpha]\!]$, $j\in [\![1,\gamma]\!]$ have
\begin{equation}
\mathbf{B}=(b_{ij})_{\alpha\times\gamma},~b_{ij}=p(\mathbf{o}_j|\mathbf{s}_i).
\end{equation}

$\pi$ is a $1\times\alpha$ probability matrix for the initial state $S_1$ when $t=1$, have
\begin{equation}
\pi=\left[~p(S_1=\mathbf{s}_1),p(S_1=\mathbf{s}_2),p(S_1=\mathbf{s}_3),...~\right]_{1\times\alpha}.
\end{equation}

Denote $\mathbb{P}_\mathrm{H}$ the HMM based CSI pattern, also as the internal Markov process of HMM. For $t=1$, $S_1=\mathbb{P}_\mathrm{H}(\pi)$; for $t>1$,
\begin{equation}\label{eq_csipattern}
S_t=\mathbb{P}_\mathrm{H}(S_1^{t-1},\mathbf{A})\overset{(a)}{=}\mathbb{P}_\mathrm{H}(S_{t-1},\mathbf{A}),~S\in \mathcal{S}.
\end{equation}
where $(a)$ is for the first-order Markov model (1-HMM) case.

Denote $\mathbb{O}_\mathrm{H}$ the HMM based CSI detecting, also as the external emission process of HMM, have
\begin{equation}\label{eq_csidetect}
O_t=\mathbb{O}_\mathrm{H}(S_t,\mathbf{B}),~O\in \mathcal{O},~S\in \mathcal{S}.
\end{equation}
\end{definition}

\begin{figure}[!h]
\centering
\includegraphics[width=9cm]{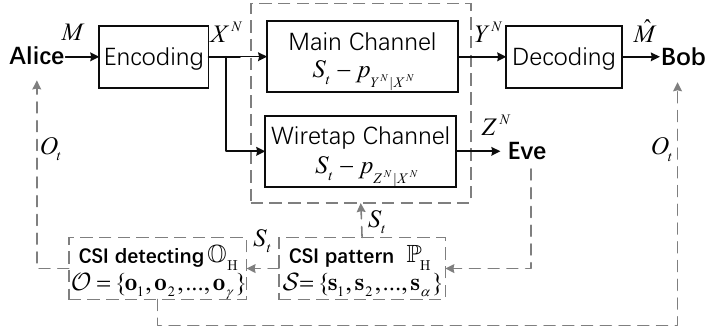}
\caption{The HMM based detectable WTC model.}
\label{fig_gvwtc}
\end{figure}

Fig.~\ref{fig_gvwtc} illustrates the communication process of the HMM based detectable WTC model, which is as follow.
\begin{itemize}
  \item[1)] Eavesdropper Eve chooses the CSI $S$ for both main channel and wiretap channel according to the CSI pattern $\mathbb{P}_\mathrm{H}$.
  \item[2)] Legitimate parties Alice and Bob detect the varying of CSI and observe the detected information $O$ according to the CSI detecting model $\mathbb{O}_\mathrm{H}$.
  \item[3)] Alice encodes the message $M$ into $N$ length codewords $X^N$ and transmits it to Bob over the main channel block with side information $O$.
  \item[4)] Bob receives $Y^N$ from the main channel block and decodes it into message $\hat{M}$ with side information $O$.
  \item[5)] Eve receives $Z^N$ form the wiretap channel block and decodes it into message $\hat{Z}^N$ according to CSI $S$.
\end{itemize}

\begin{definition}\label{def_criterion}\cite{Mahdavifar2011}
For any $(2^{NR},N)$ code over the detectable WTC model, the code performance can be measured as follow.
\begin{itemize}
\item Reliability can be measured by the bit error rate of Bob decoding the message $M$
\begin{equation}
\mathrm{P_e}=\Pr(M\neq \hat{M}).
\label{eq_errorprob}
\end{equation}
\item Security can be measured by the information leakage rate of massage $M$ to Eve
\begin{equation}
\mathrm{L_r}=\frac{I(Z^N;M)}{N}.
\label{eq_leakage}
\end{equation}
\end{itemize}
\end{definition}


\section{Active Secure Polar Coding Scheme }\label{sec_active_coding}

In this section we implement our idea of active secure coding on the HMM based detectable WTC model. Recall our idea of active secure coding solution for the uncertain CSI problem: under the detectable assumption, legitimate parties can detect the varying of hidden CSI and observe the CSI relevant information; then they can analyse the detected information and estimate the current CSI from it; according to estimated CSI, the secure scheme can actively adjust its coding strategy.

Following this idea, the framework of our active secure coding scheme over the HMM based detectable WTC model can be divided into two major parts:
\begin{itemize}
  \item A CSI learning scheme which analyses and estimates the CSI from detected information.
  \item A secure coding scheme constructed by polar codes which can actively adjust the coding strategy according to the estimated CSI.
\end{itemize}

Therefore for the rest of this section, we first present a HMM based CSI learning scheme, then we present the corresponding polar code construction, and finally we present the structure of active secure polar coding scheme by combining the CSI learning and code construction together.

\subsection{HMM Based CSI Learning Scheme}

Consider the multi-block communication over the HMM based detectable WTC model, for block $1$ to block $t$, legitimate parties can obtain the detected information sequence as $O_1^t$. With this detected information sequence, the object of the CSI learning is to estimate the current CSI $\hat{S}_t$ from the $O_1^t$. The framework of CSI learning scheme contains following three parts:
\begin{itemize}
  \item[1)] \emph{Pre-collecting stage:} initial detected information collection prior to the secure communication;
  \item[2)] \emph{CSI pattern learning:} learn the optimal parameter of the CSI pattern from the detected information sequence;
  \item[3)] \emph{CSI decoding:} decode the CSI from the detected information sequence with the learned CSI pattern.
\end{itemize}

\subsubsection{Pre-collecting stage}

We setup an $\omega$ rounds random transmission prior to the secure communication as the pre-collecting stage, in which legitimate users directly transmit random bits over the main channel, so that they can collect detected information, defined as $O_{1-\omega}^0$, from the varying CSI without any information leakage.

The purpose of this pre-collecting stage is to guarantee that legitimate parties have enough detected information (train data) for CSI pattern learning at the first few blocks of communication. For example, at the $t$-th block communication, the detected information for CSI pattern learning is $O_{1-\omega}^t$.

Also note that we use $(\omega+t)\rightarrow\infty$ to illustrate $\omega\rightarrow\infty$ with a finite $t$.

\subsubsection{CSI pattern learning}

Denote $\mathbb{L}_\mathrm{H}$ the HMM based CSI pattern learning.

Consider a $T$-times secure communication over the HMM based detectable WTC model $(\mathcal{X},\mathcal{Y},\mathcal{Z},\mathcal{S},\mathcal{O},\lambda_{\mathrm{H}})$ where $\lambda_{\mathrm{H}}=(\mathbf{A},\mathbf{B},\pi)$ is the parameter set of HMM. Then for any time $t\in[\![1,T]\!]$, the detected information obtained by legitimate parties is $O_{1-\omega}^t$.

Assuming legitimate parties know that the CSI pattern $\mathbb{P}_\mathrm{H}$ and the emission process $\mathbb{O}_\mathrm{H}$ is a HMM, but they do not know the precise parameter set $\lambda_{\mathrm{H}}$, thus for any time $t\in[\![1,T]\!]$, the CSI pattern learning is to learn the optimal estimated parameter $\hat{\lambda}_\mathrm{H}=(\mathbf{\hat{A}},\mathbf{\hat{B}},\hat{\pi})$ from the detected information sequence $O_{1-\omega}^t$, which is defined as
\begin{equation}\label{eq_csilearn}
\hat{\lambda}_\mathrm{H}=\mathbb{L}_\mathrm{H}(O_{1-\omega}^t,\mathcal{S},\mathcal{O}).
\end{equation}

Particularly for 1-HMM based detectable WTC model, we apply the \emph{Baum-Welch algorithm} \cite{Welch2003} in Appendix~\ref{sec_BHA} to implement the CSI pattern learning.

\subsubsection{CSI decoding}

Denote $\mathbb{D}_\mathrm{H}$ the HMM based CSI decoding.

Note that for any time $t\in[\![1,T]\!]$, legitimate parties obtain the detected information as $O_{1-\omega}^t$, and they have the estimated parameter set $\hat{\lambda}_\mathrm{H}=(\mathbf{\hat{A}},\mathbf{\hat{B}},\hat{\pi})$ by the CSI pattern learning.

Then the aim of  CSI decoding is to decode the optimal estimated state sequence $\hat{S}_1^t$ from the detected information $O_{1-\omega}^t$ with the learned parameter set $\hat{\lambda}_\mathrm{H}$, which is described as
\begin{equation}\label{eq_csidec}
\hat{S}_1^t=\mathbb{D}_\mathrm{H}(O_{1-\omega}^t,\hat{\lambda}_\mathrm{H},\mathcal{S},\mathcal{O}).
\end{equation}
Particularly for 1-HMM based detectable WTC model, we apply the \emph{Viterbi algorithm} \cite{Van2007} in Appendix~\ref{sec_VA} to implement this CSI decoding.

\subsection{Secure Polar Code Construction}

Next we present the construction of secure polar code for a $N$-length block with any CSI value $\mathbf{s}_i\in\mathcal{S}$. As studied in \cite{Mahdavifar2011}, the main technique for the secure polar code construction is the polarized subset division of the channel block index $[\![1,N]\!]$.

\begin{definition}(Bhattaharyya parameter)
Consider a pair of random variables $(X,Y)\sim p_{XY}$, where $X$ is a binary random variable and $Y$ is a finite-alphabet random variable. To measure the amount of randomness in $X$ given $Y$, the Bhattaharyya parameter is defined as
\begin{equation}
Z(X|Y)=2\sum_{y\in \mathcal{Y}}p_{Y}(y)\sqrt{p_{X|Y}(0|y)p_{X|Y}(1|y)}.
\label{eq_bhatdefine}
\end{equation}
\end{definition}

As we defined in Definition~\ref{def_hmm_wtc}, CSI value $\mathbf{s}_i=p_{YZ|X}^{(i)}$, $i\in[\![1,\alpha]\!]$. Assume that we know the optimal distribution of channel inputs to achieve the channel capacity under $p_{Y|X}^{(i)}$. Then for $\delta_{N}=2^{-N^\beta}$, $\mathcal{M}\rightarrow \mathcal{U}^N\rightarrow \mathcal{X}^N\rightarrow \mathcal{Y}^N,\mathcal{Z}^N$, $\beta\in\left(0,1/2\right)$, according to the source polarization theory\cite{Arikan2010} and channel polarization theory\cite{Arikan2009}, we can have the following polarized result of CSI value $\mathbf{s}_i$:

\begin{itemize}
\item Source polarization
\begin{equation}
\begin{split}
&\mathcal{H}_X^{(i)}=\left\{j\in[\![1,N]\!]:Z\left(U_j|U^{j-1}_1\right)\geq1-\delta_N \right\},\\
&\mathcal{L}_X^{(i)}=\left\{j\in[\![1,N]\!]:Z\left(U_j|U^{j-1}_1\right)\leq\delta_N \right\}.
\end{split}
\end{equation}
\item Main channel polarization
\begin{equation}
\begin{split}
&\mathcal{H}_{X|Y}^{(i)}=\left\{j\in[\![1,N]\!]:Z\left(U_j|U^{j-1}_1,Y^N\right)\geq1-\delta_N \right\},\\
&\mathcal{L}_{X|Y}^{(i)}=\left\{j\in[\![1,N]\!]:Z\left(U_j|U^{j-1}_1,Y^N\right)\leq\delta_N \right\}.
\end{split}
\end{equation}
\item Wiretap channel polarization
\begin{equation}
\begin{split}
&\mathcal{H}_{X|Z}^{(i)}=\left\{j\in[\![1,N]\!]:Z\left(U_j|U^{j-1}_1,Z^N\right)\geq1-\delta_N \right\},\\
&\mathcal{L}_{X|Z}^{(i)}=\left\{j\in[\![1,N]\!]:Z\left(U_j|U^{j-1}_1,Z^N\right)\leq\delta_N \right\}.
\end{split}
\end{equation}
\end{itemize}

Based on above polarized results, we divide the index $[\![1,N]\!]$ as follow, which is similar as the structure in \cite{Wei2015,Gulcu2015}.
\begin{equation}\label{eq_division1}
\begin{split}
&\mathcal{I}^{(i)}=\mathcal{H}_X^{(i)}\cap \mathcal{L}_{X|Y}^{(i)}\cap \mathcal{H}_{X|Z}^{(i)},\\
&\mathcal{F}^{(i)}=\mathcal{H}_X^{(i)}\cap \left(\mathcal{L}_{X|Y}^{(i)}\right)^c\cap \mathcal{H}_{X|Z}^{(i)},\\
&\mathcal{R}^{(i)}=\mathcal{H}_X^{(i)}\cap \mathcal{L}_{X|Y}^{(i)}\cap \left(\mathcal{H}_{X|Z}^{(i)}\right)^c,\\
&\mathcal{B}^{(i)}=\mathcal{H}_X^{(i)}\cap \left(\mathcal{L}_{X|Y}^{(i)}\right)^c\cap \left(\mathcal{H}_{X|Z}^{(i)}\right)^c,\\
&\mathcal{D}^{(i)}=\left(\mathcal{H}_X^{(i)}\right)^c.\\
\end{split}
\end{equation}

Since legitimate parties cannot know the CSI of next block, in order to implement the multi-block chaining structure\cite{Mahdavifar2011}, we have to assume that $|\mathcal{I}^{(i)}|>\max\limits_\mathcal{S} |\mathcal{B}|$, and then construct the subset $\mathcal{E}^{(i)}$ as follow
\begin{equation}\label{eq_division2}
\mathcal{E}^{(i)}\subset\mathcal{I}^{(i)},~|\mathcal{E}^{(i)}|=\max_\mathcal{S} |\mathcal{B}|.
\end{equation}

Finally for every $\mathbf{s}_i=p_{YZ|X}^{(i)}$, we have the divided subsets $\mathcal{I}^{(i)}\setminus\mathcal{E}^{(i)}$, $\mathcal{E}^{(i)}$, $\mathcal{F}^{(i)}$, $\mathcal{R}^{(i)}$, $\mathcal{B}^{(i)}$ and $\mathcal{D}^{(i)})$ for the secure polar code.

Then the functions of these subsets are as follow: subset $\mathcal{I}^{(i)}\setminus\mathcal{E}^{(i)}$ is secure and reliable, which is for information bits; subset $\mathcal{E}^{(i)}$, with a fixed size for all $\mathbf{s}_i$, is secure and reliable, which is for functional random bits; subset $\mathcal{F}^{(i)}$ is secure but unreliable, which is for frozen bits; subset $\mathcal{R}^{(i)}$ is reliable but insecure, which is for uniformly distributed random bits; subset $\mathcal{B}^{(i)}$ is neither secure nor reliable, thus which is constructed to retransmit the random bits in $\mathcal{E}^{(i)}$ of previous round; subset $\mathcal{D}^{(i)}$ is for deterministic bits calculated by
\begin{equation}
u_j=\arg \max_{u\in\{0,1\}}p_{U_j|U_1^{j-1}}(u|u_1^{j-1}),~j\in\mathcal{D}^{(i)}.
\label{eq_determinbit}
\end{equation}

\subsection{Active secure polar coding scheme}

Next we present the active secure polar coding scheme by combining the HMM based CSI learning scheme and secure polar code.

\begin{figure}[!h]
\centering
\includegraphics[width=9cm]{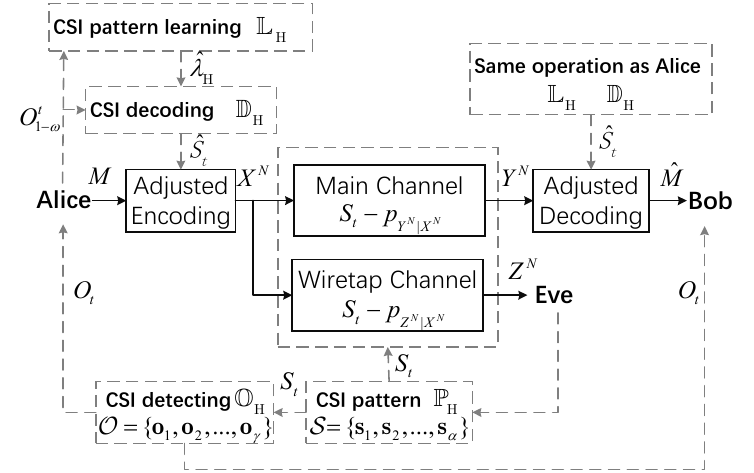}
\caption{The framework of the active secure coding scheme.}
\label{fig_wtccoding}
\end{figure}

\begin{figure}[!h]
\centering
\includegraphics[width=9cm]{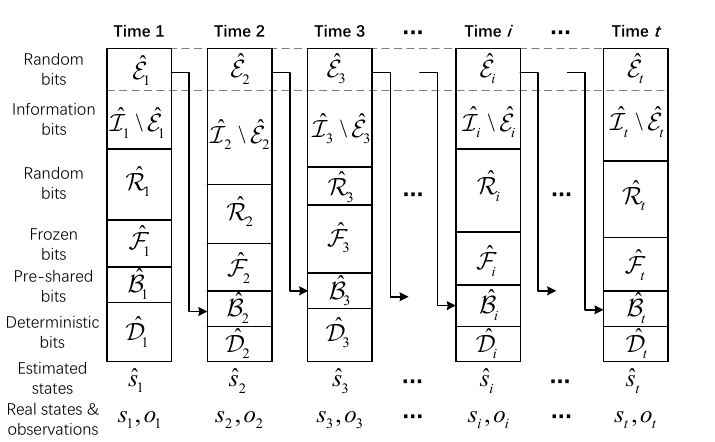}
\caption{The construction of secure polar code with estimated CSI.}
\label{fig_blocks}
\end{figure}

The framework of the active secure polar coding scheme is illustrated in Fig.~\ref{fig_blocks}. Every time when eavesdropper Eve changes the CSI $S_t$ according to the CSI pattern $\mathbb{P}_\mathrm{H}$\eqref{eq_csipattern}, legitimate parties, Alice and Bob, observes the detected information as $O_{1-\omega}^t$ by the CSI detecting $\mathbb{O}_\mathrm{H}$\eqref{eq_csidetect}. Then by CSI pattern learning $\mathbb{L}_\mathrm{H}$\eqref{eq_csilearn} and the CSI decoding $\mathbb{D}_\mathrm{H}$ \eqref{eq_csidec}, the current CSI can be estimated from the detected information as $\hat{S}_t$. Therefore, as illustrated in Fig.~\ref{fig_blocks}, with the estimated CSI $\hat{S}_t$ legitimate parties can perform the polarized subset division \eqref{eq_division1}\eqref{eq_division2} of the secure polar code and then establish the multi-block chaining structure for secure and reliable  communication.

Now we present the active secure polar coding scheme in detail. Consider a $T$ blocks secure communication from time $1$ to time $T$ over the HMM based detectable WTC model $(\mathcal{X},\mathcal{Y},\mathcal{Z},\mathcal{S},\mathcal{O},\mathbf{A},\mathbf{B},\pi)$ with pre-collected information $O_{1-\omega}^0$ and confidential message $M_1^T\in\mathcal{M}$. Assume that legitimate parties know the full set of $\mathcal{S},\mathcal{O}$ and the optimal channel input distribution for each CSI value. Then the structure of active secure polar coding scheme is as follow.

\begin{itemize}
\item[\textbf{\emph{1)}}] \textbf{\emph{CSI learning:}} for $t$-th time, legitimate parties have the detected information sequence as $o_{1-\omega}^t$, then they learn the CSI pattern from $o_{1-\omega}^t$ by
\begin{equation}
\hat{\lambda}_\mathrm{H}=\mathbb{L}_\mathrm{H}(o_{1-\omega}^t,\mathcal{S},\mathcal{O}),
\end{equation}
 and decode the estimated CSI sequence $\hat{s}_1^t$ from $o_{1-\omega}^t$ by
\begin{equation}
\hat{s}_1^t=\mathbb{D}_\mathrm{H}(o_{1-\omega}^t,\hat{\lambda}_\mathrm{H},\mathcal{S},\mathcal{O}).
\end{equation}
\item[\textbf{\emph{2)}}] \textbf{\emph{Polarized subsets division:}} for $t$-th time, based on the estimated CSI $\hat{s}_t$, perform the polar subsets division of channel index $N$ to obtain $(\mathcal{\hat{I}}_t,\mathcal{\hat{F}}_t,\mathcal{\hat{R}}_t,\mathcal{\hat{B}}_t,\mathcal{\hat{D}}_t,\mathcal{\hat{E}}_t)$.
\item[\textbf{\emph{3)}}] \textbf{\emph{Encoding:}} for $t$-th time, assign the $u^N$ as follow:
\begin{itemize}
\item[-] $u^{\mathcal{\hat{I}}_t\setminus\mathcal{\hat{E}}_t}$: assigned with information bits of $M_t$;
\item[-] $u^{\mathcal{\hat{F}}_t}$: assigned with frozen bits;
\item[-] $u^{\mathcal{\hat{R}}_t}$: assigned with uniformly distributed random bits;
\item[-] $u^{\mathcal{\hat{E}}_t}$: assigned with uniformly distributed random bits;
\item[-] $u^{\mathcal{\hat{D}}_t}$: assigned with deterministic bits calculated by \eqref{eq_determinbit};
\item[-] $u^{\mathcal{\hat{B}}_t}$: if $t=1$, assigned with a pre-shared random bits; if $t>1$, assigned with the first $|\mathcal{\hat{B}}_t|$ bits of $u^{\mathcal{\hat{E}}_{t-1}}$ in time $t-1$.
\end{itemize}
Then encode $u^N$ into the optimally distributed channel input $x^N$ by polar encoding $x^N=u^N\mathbf{G}_N$, and transmit $x^N$ over the main channel block.
\item[\textbf{\emph{4)}}] \textbf{\emph{Decoding:}} for $t$-th time, legitimate user Bob receives $y^N$ and decodes it into the estimated $\hat{u}^N$ with the estimated CSI $\hat{s}_t$ by the succussive cancelation decoding \cite{Arikan2009}.
\begin{itemize}
\item[-] for $j\in \mathcal{\hat{I}}_t\cup\mathcal{\hat{R}}_t$,
\begin{equation}
\hat{u}_j=\arg \max \limits_{u\in\left\{0,1\right\}}p_{U_j|U_1^{j-1}Y^N}\left(u|\hat{u}_1^{j-1}y^N\right)
\end{equation}
\item[-] for $j\in\mathcal{\hat{F}}_t$, $\hat{u}_j$ is directly decoded as the frozen bits;
\item[-] for $j\in\mathcal{\hat{B}}_t$, if $t=1$, $\hat{u}_j$ is directly decoded as the pre-shared bits, if $t>1$, $\hat{u}_j$ is directly decoded as the correspondent bit of $\hat{u}^{\mathcal{\hat{E}}_{i-1}}$ in time $i-1$;
\item[-] for $t\in\mathcal{\hat{D}}_t$,
\begin{equation}
\hat{u}_j=\arg \max \limits_{u\in\left\{0,1\right\}}p_{U_j|U_1^{j-1}}\left(u|\hat{u}_1^{j-1}\right)
\end{equation}
\end{itemize}
\end{itemize}

\section{Simulations}\label{sec_simulation}

In this section, we test the performance the active secure polar coding scheme. Particularly, we build a concrete 1-HMM based detected WTC model for the simulation.
\begin{itemize}
  \item For CSI pattern: let $\mathbf{A}=[0.95,0.05;0.10,0.90]$, $\pi=[0.95,0.05]$, and CSI uncertain set $\mathcal{S}=\{(0.2,0.5),(0.3,0.6)\}$, where each CSI $\mathbf{s}_i=(\epsilon_m,\epsilon_w)$ means a BEC pair with erase probabilities $\epsilon_m$ and $\epsilon_w$ respectively for the main channel and wiretap channel.
  \item For CSI detecting: let $\mathbf{B}=[1/6,1/6,1/6,1/6,1/6,1/6;1/10,$ $1/10,1/10,1/10,1/10,1/2]$, and $O=\{1,2,3,4,5,6\}$ that represents the $6$ possible values of detected information.
\end{itemize}

\subsection{Performance of CSI Learning Scheme}

First we test the performance of the HMM based CSI learning scheme, including both HMM based CSI pattern learning and CSI decoding.

%

To run the simulation for the CSI pattern learning, we setup an initial estimated parameter set $\hat{\lambda}_\mathrm{H}$ as $\hat{\mathbf{A}}=[0.80,0.20;0.20,0.80]$, $\hat{\mathbf{B}}=[1/5, 1/5, 1/5, 1/5, 1/10, 1/10; 1/8, 1/8, 1/8, 1/8, 1/4, 1/4]$ and $\hat{\pi}=[0.80,0.20]$. Then the CSI pattern learning is the updating process of this estimated parameter set $\hat{\lambda}_\mathrm{H}$ based on the detected information $O_{1-\omega}^t$. We use the \emph{Euclidean Distance} between $\lambda_\mathrm{H}$ and $\hat{\lambda}_\mathrm{H}$ to measure the accuracy of the estimation, which is denoted as $\|\lambda_\mathrm{H}-\hat{\lambda}_\mathrm{H} \|$.

\begin{figure}[!h]
\centering
\includegraphics[width=8.5cm]{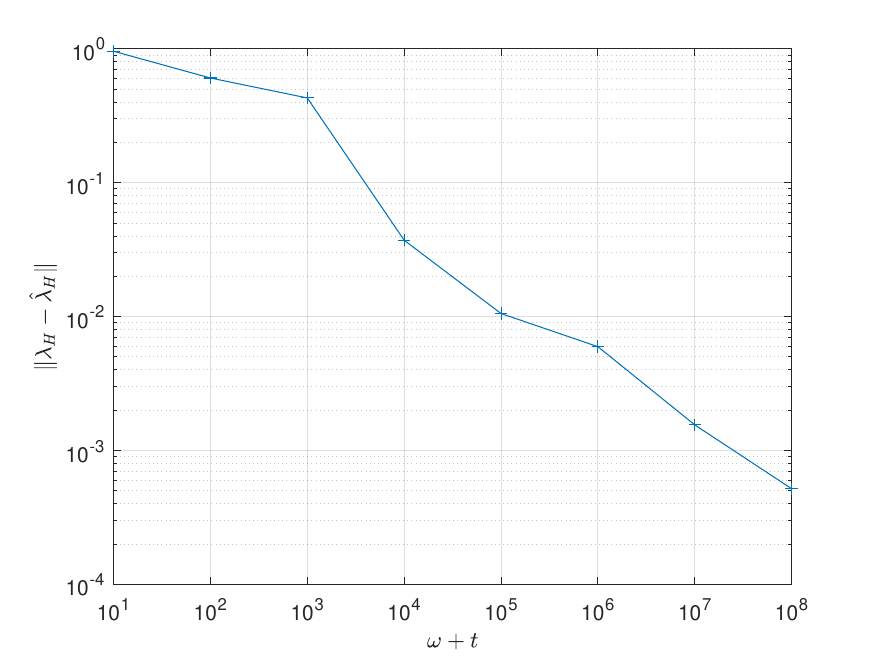}
\caption{The distance between $\lambda_\mathrm{H}$ and $\hat{\lambda}_\mathrm{H}$.}
\label{fig_stim_learn}
\end{figure}

The simulation result of the CSI pattern learning is illustrated in Fig.~\ref{fig_stim_learn}. We can observe that when the length of detected information $O_{1-\omega}^t$ increasing, the estimated parameter set $\hat{\lambda}_\mathrm{H}$ is getting closer to the actual parameter set $\lambda_\mathrm{H}$, which indicates the estimation of $\lambda_\mathrm{H}$ can reach an acceptable level of accuracy with enough detected information.

Then we run the simulation for the CSI decoding based on the estimated parameter set $\hat{\lambda}_\mathrm{H}$ from the CSI learning simulation, and analyze the CSI error rate of decoding $\hat{S}_{1-\omega}^t$ from detected information $O_{1-\omega}^t$.

\begin{figure}[!h]
\centering
\includegraphics[width=8.5cm]{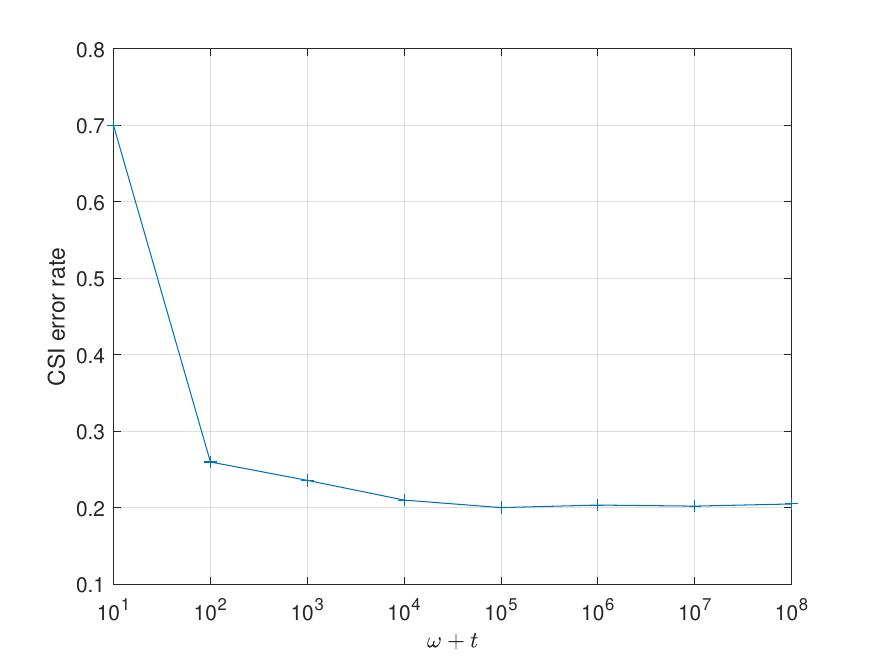}
\caption{The CSI error rate.}
\label{fig_stim_viterbi}
\end{figure}

The simulation result of the CSI decoding is illustrated in Fig.~\ref{fig_stim_viterbi}. We can observe that with the increasing of detected information $O_{1-\omega}^t$, the error rate of CSI decoding is decreasing and but then remains on a relatively low level.  Note that for the CSI learning scheme, the CSI $S$ is coded into detected information $O$ and then decoded into estimated CSI $\hat{S}$ which forms a Markov chain $S\rightarrow O \rightarrow \hat{S}$ that takes values in $\mathcal{S} \rightarrow \mathcal{O}\rightarrow \mathcal{S}$. The CSI error rate can be defined as
\begin{equation}
  \mathrm{P_{e\_csi}}=\Pr(S\neq\hat{S}).
\end{equation}
Then according to the \emph{Fano's inequation}, $\mathrm{P_{e\_csi}}$ satisfies
\begin{equation}\label{eq_fano}
  H(S|O)\leq H(\mathrm{P_{e\_csi}})+\mathrm{P_{e\_csi}}\log(|\mathcal{S}|-1).
\end{equation}
Thus there is a lower bound of the CSI error rate for the CSI decoding.



\subsection{Performance of Active Secure Polar Coding Scheme}

Next we analyze the performance of the active secure polar coding scheme, including both reliability and security.

Let $M$ be the binary confidential message which is uniformly distributed on $\{0,1\}$. Let $\omega=2000$ be the rounds of pre-collecting stage and $T=8000$ be the number of channel blocks for the secure communication. Let $\beta=0.25$ for the polarized subset division of secure polar codes. Then we run the simulation of active secure polar coding with $n=7,8,9,10$ successively.

\begin{figure}[!h]
\centering
\includegraphics[width=8.5cm]{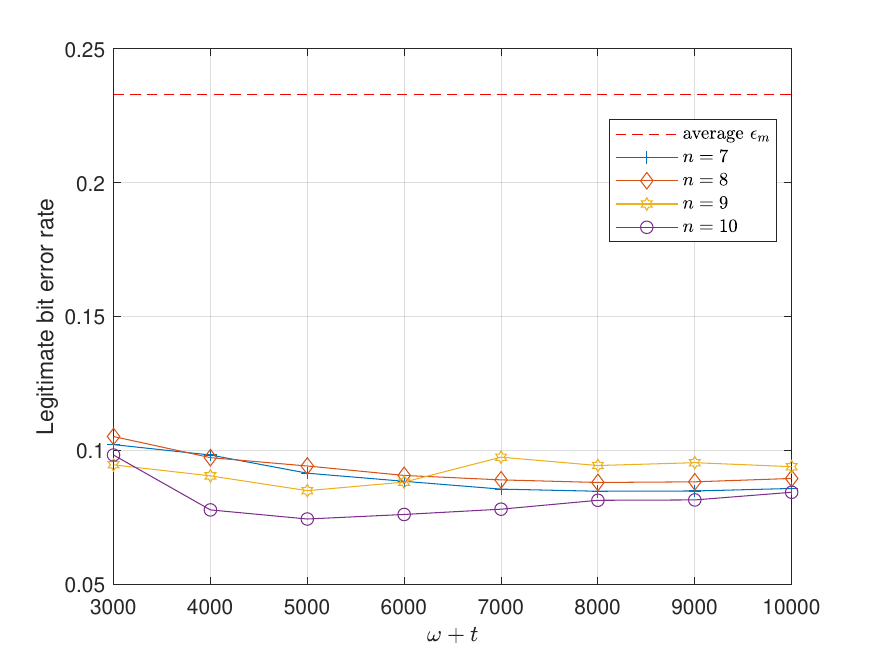}
\caption{Bit error rate of message $M_1^t$ for Bob.}
\label{fig_stim_legitmate_ber}
\end{figure}

Fig.~\ref{fig_stim_legitmate_ber} illustrates the bit error rate of legitimate parties decoding the confidential message $M_1^t$. We can observe that with the increasing of the block number $\omega+t$, the bit error rate drops below the $0.1$ and then remains on a relatively low level comparing with the average main channel erase probability, which indicates that an accepted level of reliability can be achieved.

But on the other hand, with the increasing of block length $N=2^n$, there is no obvious vanishing trend for the bit error rate, thus the perfect reliability criterion $\lim_{N\rightarrow\infty}\mathrm{P_e}=0$ cannot be achieved, and the main reason for this failure is the existence of lower bound for the CSI error rate when estimating the CSI from the detected information.

\begin{figure}[!h]
\centering
\includegraphics[width=8.5cm]{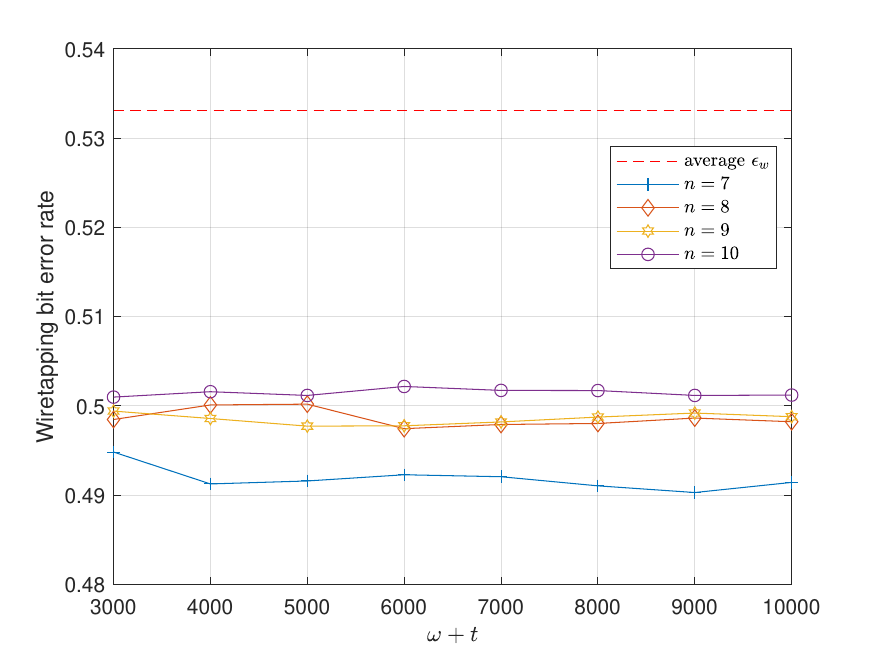}
\caption{Bit error rate of message $M_1^t$ for Eve.}
\label{fig_stim_wiretap_ber}
\end{figure}

\begin{figure}[!h]
\centering
\includegraphics[width=8.5cm]{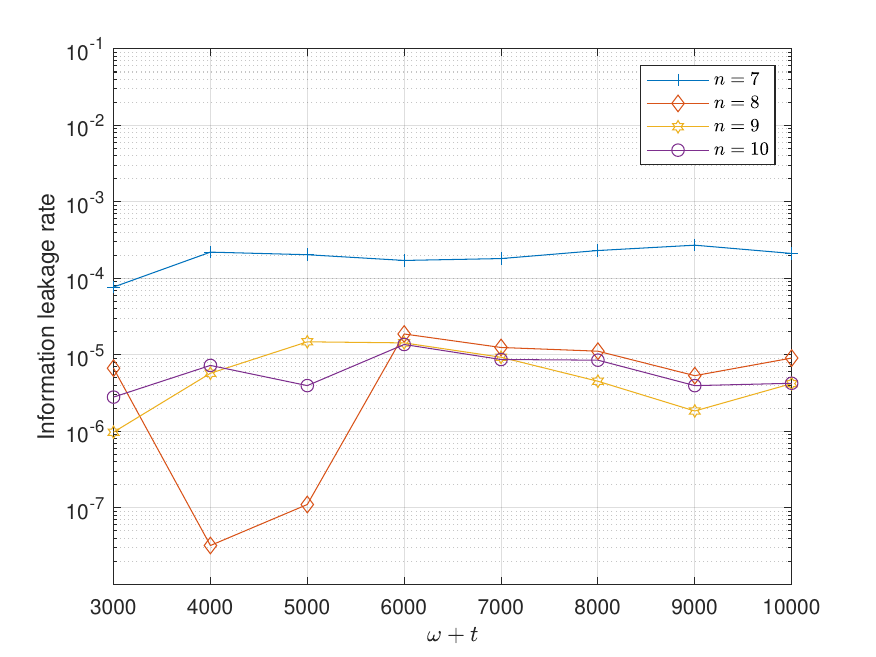}
\caption{Information leakage rate of message $M_1^t$ for Eve.}
\label{fig_stim_leakge}
\end{figure}

Fig.~\ref{fig_stim_wiretap_ber} illustrates the bit error rate of eavesdropper decoding the confidential message $M_1^t$, and the corresponding information leakage rate of $M_1^t$ is illustrated in Fig.~\ref{fig_stim_leakge}. From Fig.~\ref{fig_stim_wiretap_ber}, we can observe that the bit error rate of eavesdropper decoding the message $M_1^t$ remains close to the $0.5$ when the block number $\omega+t$ is increasing. Note that message $M$ is uniformly distributed over $\{0,1\}$, if the bit error rate gets closer to $0.5$, the information leakage rate will get lower. Thus in Fig.~\ref{fig_stim_wiretap_ber} the information leakage rate remains at a low level which indicates that an acceptable level of security can be achieved.

But also because of the lower bound for the CSI error rate, there is no trend for bit error rate approaching the $0.5$ or information leakage rate vanishing with an increasing block length $N=2^n$. Thus the perfect security criterion $\lim_{N\rightarrow\infty}\mathrm{L_r}=0$ also cannot be achieved.

\section{Conclusion}\label{sec_con}

In this paper, we have proposed a new solution for the uncertain CSI problem called active secure coding, which combines the machine learning methods with the traditional physical layer secure coding scheme to achieve secure and reliable communication.

First, we use HMM to model the CSI pattern and CSI detecting, that the varying of channel block CSI is a Markov process, and the detected information is a stochastic emission from the current CSI. Then we combine the HMM with the compound WTC model to build the HMM based detectable WTC model. Next, we present a CSI learning scheme to learn the CSI from the detected information by applying the Baum-Welch algorithm and the Viterbi algorithm. Besides, we setup a pre-collecting stage to collect training data prior to the secure communication. Further, we construct an explicit secure polar codes based on the learned CSI, and combine it with the CSI learning scheme to form our active secure polar coding scheme.

At last, we carry out simulations to test the performance of the active secure polar coding scheme. For the CSI learning scheme, simulation results show that the parameter $\lambda$ can be correctly learned from the detected information, but the lower bound of CSI error rate exists for estimating the CSI from detected information. Because of this lower bound, the secure coding scheme can achieve an acceptable level of reliability and security (low bit error rate and information leakage rate), but fails to achieves perfect reliability or perfect security.

Our future work will focus on the remaining problem of achieving perfect reliability and security. Particularly, instead of decoding CSI from detected information, we will try to construct secure polar codes only with the correctly learned parameter $\lambda$.

\section*{Acknowledgment}

This work is supported in part by the Natural Science Foundation of Hubei Province (Grant No.2019CFB137) and the Fundamental Research Funds for the Central Universities (Grant No.2662017QD042, No.2662018JC007).

%
%
%

\ifCLASSOPTIONcaptionsoff
  \newpage
\fi




\begin{thebibliography}{1}

\bibitem{Wyner1975}
Wyner A. D.: `The wire-tap channel', \textit{Bell System Tech. J.}, 1975, \textbf{54}, pp. 1355--1387

\bibitem{Arikan2009}
Ar{\i}kan E.: `Channel polarization: a method for constructing capacity achieving codes for symmetric binary-input memoryless channels', \textit{IEEE Trans. Inf. Theory}, 2009, \textbf{55}, pp. 3051--3073

\bibitem{Mahdavifar2011}
Mahdavifar H., Vardy A.: `Achieving the secrecy capacity of Wiretap channels using Polar codes', \textit{IEEE Trans. Inf. Theory}, 2011, \textbf{57}, pp. 6428--6443

\bibitem{Vard2013strong}
\c{S}a\c{s}o\u{g}lu E., Vardy A.: `A New Polar Coding Scheme for Strong Security on Wiretap Channels', \textit{Proc. IEEE Int. Symp. Inf. Theory (ISIT)}, 2013, pp. 1117--1121

\bibitem{Hassani2014}
Hassani S. H., Urbanke R.: `Universal polar code', \textit{Proc. IEEE Int. Symp. Inf. Theory (ISIT)}, 2014, pp. 1451--1455

\bibitem{Wei2015}
Wei Y.-P., Ulukus S.: `Polar coding for the general wiretap channel', \textit{Proc. IEEE Int. Theory Workshop}, 2015, pp. 1--5

\bibitem{Gulcu2015}
Gulcu T. C., Barg A.: `Achieving secrecy capacity of the wiretap channel and broadcast channel with a confidential component', \textit{Proc. IEEE Inf. Theory Workshop}, 2015, pp. 1--5

\bibitem{Zheng2017}
Zheng M., Tao M., Chen W., Ling C.: `Secure Polar Coding for the Two-Way Wiretap Channel', \textit{IEEE Access}, 2018, pp. 1--1.

\bibitem{Ozarow1984}
Ozarow L. H., Wyner A. D.: `Wire-tap channel II', \textit{AT\&T Bell Laboratories technical journal}, 1984, \textbf{63}, pp. 2135--2157

\bibitem{Wang2013}
Wang P., Safavi-Naini R.: `A model for adversarial wiretap channels', \textit{IEEE Trans. Inf. Theory}, 2013, \textbf{62}, pp. 970--983

\bibitem{Goldfeld2016}
Goldfeld Z., Cuff P., Permuter H. H.: `Arbitrarily varying wiretap channels with type constrained states', \textit{IEEE Trans. Inf. Theory}, 2016, \textbf{62}, pp. 7216--7244

\bibitem{Schaefer2015}
Schaefer R. F., Boche H., Poor H. V.: `Secure communication under channel uncertainty and adversarial attacks', \textit{Proc. of the IEEE}, 2015, \textbf{103}, pp. 1796--1813

\bibitem{Cheng2012}
Cheng F., Yeung R. W., Shum K. W.: `Imperfect secrecy in wiretap channel II', \textit{Proc. IEEE Int. Symp. Inf. Theory (ISIT)}, 2012, pp. 71--75

\bibitem{Nafea2016}
Nafea M., Yener A.: `A new wiretap channel model and its strong secrecy capacity', \textit{Proc. IEEE Int. Symp. Inf. Theory (ISIT)}, 2016, pp. 2804--2808

\bibitem{Tahmasbi2017}
Tahmasbi M., Bloch M. R., Yener A.: `Learning adversary's actions for secret communication', \textit{Proc. IEEE Int. Symp. Inf. Theory (ISIT)}, 2017, pp. 2708--2712

\bibitem{Tahmasbi2019}
Tahmasbi M., Bloch M. R., Yener A.: `Learning an adversary's actions for secret communication', \textit{Online arXiv:1807.08670v2 [cs.IT]}, 2019

\bibitem{Dai2017}
Dai B., Ma Z., Luo Y.: `Finite state markov wiretap channel with delayed feedback', \textit{IEEE Transactions on Information Forensics and Security}, 2017, pp. 746--760

\bibitem{Wang2018}
Wang T., Wen C. K., Jin S. and Li G. Y.: `Deep learning-based CSI feedback approach for time-varying massive MIMO channels', \textit{IEEE Wireless Commun. Lett.}, 2018, pp. 1--1

\bibitem{Arikan2010}
Ar{\i}kan E.: `Source polarization', \textit{IEEE Int. Symp. Inf. Theory (ISIT)}, 2010, pp. 899--903

\bibitem{Welch2003}
Welch L. R.: `Hidden Markov Models and the Baum-Welch algorithm', \textit{IEEE Information Theory Society Newsletter}, 2003, \textbf{53}, pp. 194--211

\bibitem{Van2007}
Van B. L., Garcia-Salicetti S., Dorizzi B.: `On using the viterbi path along with HMM likelihood information for online signature verification', \textit{IEEE Transactions on Systems Man \& Cybernetics Part B}, 2007, \textbf{37}, pp. 1237--1247

\bibitem{Shor1994}
Shor P. W.: `Algorithms for quantum computation: discrete logarithms and factoring', \textit{Foundations of Computer Science, 1994 Proceedings. Symposium on IEEE}, 2002, pp. 124--134

\bibitem{Monz2016}
Monz T., Nigg D., Martinez E. A., Brandl M. F., Schindler P., Rines R., Wang S. X., Chuang I. L., Blatt R.: `Realization of a scalable shor algorithm', \textit{Science}, 2016, \textbf{351}, 1068--1070


\end{thebibliography}
%

\section{Appendices}

\subsection{The Baum-Welch Algorithm for CSI Pattern Learning}\label{sec_BHA}
Considering the HMM model $(\mathcal{S},\mathcal{O},\mathbf{A},\mathbf{B},\pi)$ introduced in Section~\ref{sec_HMMwtc} with unknown CSI time sequence $s_1^t$ and detected observation time sequence $o_1^t$. Assume we know the full set of $\mathcal{S}$ and $\mathcal{O}$. Define $b_i(o_j)=p(o_j|\mathbf{s}_i)$ and set random initial conditions to $\lambda_\mathrm{H}=(\mathbf{A},\mathbf{B},\pi)$. Then the Baum-Welch algorithm \cite{Welch2003} for CSI pattern learning is as follow.

\begin{itemize}
\item[i.] Forward procedure: for $i\in[\![1,\alpha]\!]$, $k\in [\![1,t]\!]$, denote
\begin{equation}
\mu_i(k)=p(o_1^k,S_k=\mathbf{s}_i|\lambda_\mathrm{H})
\end{equation}

initialization
\begin{equation}
\mu_i(1)=\pi_i b_i(o_1)
\end{equation}

recursion
\begin{equation}
\mu_i(k+1)=b_j(o_{k+1})\sum_{j=1}^\alpha \mu_j(k)a_{ji}
\end{equation}

\item[ii.] Backward procedure: for $i\in[\![1,\alpha]\!]$, $k\in [\![1,t]\!]$, denote
\begin{equation}
\rho_i(k)=p(o_1^k|S_k=\mathbf{s}_i,\lambda_\mathrm{H})
\end{equation}

initialization
\begin{equation}
\rho_i(t)=1
\end{equation}

recursion
\begin{equation}
\rho_i(k)=\sum_{j=1}^\alpha \rho_j(k+1)a_{ij}b_j(o_{k+1})
\end{equation}

\item[iii.] Update: denote $\upsilon_i(k)=p(S_k=\mathbf{s}_i|o_1^k,\lambda_\mathrm{H})$ and $\phi_{ij}(k)=p(S_k=\mathbf{s}_i,S_{k+1}=\mathbf{s}_j|o_1^k,\lambda_\mathrm{H})$. Then according to the Bayes' theorem, have
\begin{equation}
\upsilon_i(k)=\frac{p(S_k=\mathbf{s}_i,o_1^k|\lambda_\mathrm{H})}{p(o_1^k|\lambda_\mathrm{H})}=\frac{\mu_i(k)\rho_i(k)}{\sum_{j=1}^\alpha\mu_j(k)\rho_j(k)}
\end{equation}
\begin{equation}
\begin{split}
\phi_{ij}(k)&=\frac{p(S_k=\mathbf{s}_i,S_{k+1}=\mathbf{s}_j,o_1^k|\lambda_\mathrm{H})}{p(o_1^k|\lambda_\mathrm{H})}\\
=&\frac{\mu_i(k)a_{ij}\rho_j(k+1)b_j(o_{k+1})}{\sum_{i=1}^\alpha\sum_{j=1}^\alpha \mu_i(k)a_{ij}\rho_j(k+1)b_j(o_{k+1})}
\end{split}
\end{equation}

Then the parameter set $\lambda_\mathrm{H}$ of the HMM can be updated by
\begin{equation}
\pi^*_i=\upsilon_i(1)
\end{equation}
\begin{equation}
a^*_{ij}=\frac{\sum_{k=1}^{t-1}\phi_{ij}(k)}{\sum_{k=1}^{t-1}\upsilon_i(k)}
\end{equation}
\begin{equation}
b^*_{ij}=\frac{\sum_{k+1}^t \mathbbm{1}_{o_k=\mathbf{o}_j}\upsilon_i(k)}{\sum_{k+1}^t\upsilon_i(k)}
\end{equation}
where $i\in[\![1,\alpha]\!]$,  $j\in[\![1,\gamma]\!]$, $k\in [\![1,t]\!]$
\begin{equation}
\mathbbm{1}_{o_k=\mathbf{o}_j}=
\begin{cases}
1~\text{if}~o_k=\mathbf{o}_j\\
0~\text{otherwise}
\end{cases}
\end{equation}

\end{itemize}

\subsection{The Viterbi Algorithm for CSI Decoding}\label{sec_VA}

Following the notation for HMM in Appendix~\ref{sec_BHA}, assume that we know the full set of $\mathcal{S}$, $\mathcal{O}$ and the parameter set $\lambda_\mathrm{H}$ and have the observation $o_1^t$.

For $i\in[\![1,\alpha]\!]$, $k\in [\![1,t]\!]$, denote $\kappa_i(t)$ as the probability of the most likely path $\hat{s}_1^k$ with $\hat{s}_t=\mathbf{s}_i$ that generates the observation $o_1^k$, denote $\tau_i(k)$ as the $\hat{s}_{k-1}$ of the most likely path $\hat{s}_1^{k-1}$ with $\hat{s}_k=\mathbf{s}_i$, have
\begin{equation}
\kappa_i(k)=\max_{s\in\mathcal{S}}p(\hat{s}_1^k,\hat{s}_t=\mathbf{s}_i,o_1^k,\lambda_\mathrm{H})
\end{equation}

Then the Viterbi algorithm \cite{Van2007} for CSI decoding is as follow.

\begin{itemize}

\item[i.] Initialization with
\begin{equation}
\begin{split}
\kappa_i(1)&=\pi_ib_i(o_1)\\
\tau_i(1)&=0
\end{split}
\end{equation}

\item[ii.] Recursion for $i\in[\![1,\alpha]\!]$, $k\in [\![2,t]\!]$
\begin{equation}
\begin{split}
\kappa_i(k)&=\max_{i\in[\![1,\alpha]\!]}\left[\kappa_i(k-1)a_{ij}b_j(o_k)\right]\\
\tau_i(k)&=\arg \max_{i\in[\![1,\alpha]\!]}\left[\kappa_i(k-1)a_{ij}b_j(o_k)\right]
\end{split}
\end{equation}

\item[iii.] End for
\begin{equation}
\begin{split}
p^*&=\max_{i\in[\![1,\alpha]\!]}\kappa_i(t)\\
s^*_t&=\arg \max_{i\in[\![1,\alpha]\!]}\kappa_i(t)
\end{split}
\end{equation}

\item[vi.] Path trace, for $k=t-1,t-2,...,1$
\begin{equation}
s^*_{k}=\tau_{s^*_{k+1}}(k+1)
\end{equation}

\end{itemize}

%

\begin{IEEEbiographynophoto}{Yizhi Zhao}
received his Ph.D. degree in the School of Optical and Electronic Information from the Huazhong University of Science and Technology, Wuhan,
China, in 2017.

He is currently an Assistant Professor with the College of Informatics, Huazhong Agricultural University. His research interests include physical layer coding, VLSI design and machine learning.
\end{IEEEbiographynophoto}

%
%




\end{document}